\documentclass{elsart}

\usepackage{amsmath,amssymb,graphicx,bm}

\newcommand{\beqn}{\begin{equation}}
\newcommand{\eeqn}{\end{equation}}
\newcommand{\bea}{\begin{eqnarray}}
\newcommand{\eea}{\end{eqnarray}}
\newcommand{\ba}{\begin{align}}
\newcommand{\ea}{\end{align}}
\newcommand{\vlowk}{V_{{\rm low}\,k}}

\newcommand{\fmi}{\, \text{fm}^{-1}}
\newcommand{\mev}{\, \text{MeV}}

\begin{document}

\begin{frontmatter}

\title{Convergence of the Born Series \\ 
with Low-Momentum Interactions}

\author{S.K.\ Bogner}$^1$,
\ead{bogner@mps.ohio-state.edu}
\author{R.J.\ Furnstahl}$^1$,
\ead{furnstahl.1@osu.edu}
\author{S.\ Ramanan}$^1$ and
\ead{suna@mps.ohio-state.edu}
\author{A.\ Schwenk}$^2$
\ead{schwenk@u.washington.edu}
\address{$^1$Department of Physics,
The Ohio State University, Columbus, OH\ 43210\\
$^2$Department of Physics,
University of Washington, Seattle, WA\ 98195-1560}


\begin{abstract}
The nonperturbative nature of nucleon-nucleon interactions
as a function of a momentum cutoff is studied using Weinberg eigenvalues
as a diagnostic.
This investigation extends 
an earlier study of the perturbative convergence of the
Born series to partial waves beyond
the $^3$S$_1$--$^3$D$_1$ channel and to positive energies.  
As the cutoff is lowered using renormalization-group or model-space 
techniques, the evolution of nonperturbative features at large cutoffs 
from strong short-range repulsion and the iterated tensor interaction are
monitored via the complex Weinberg eigenvalues.
When all eigenvalues lie within the unit circle, 
the expansion of the scattering amplitude in terms of the interaction 
is perturbative,
with the magnitude of the largest eigenvalue setting the rate of
convergence.
Major decreases in the magnitudes of repulsive eigenvalues
are observed as the Argonne $v_{18}$, CD-Bonn or Nijmegen potentials are 
evolved to low momentum,  even though two-body
observables are unchanged.
For chiral EFT potentials, running the cutoff lower tames the impact
of the tensor force and of new nonperturbative features entering at
N$^3$LO.
The efficacy of separable approximations to nuclear interactions
derived from the Weinberg analysis 
is studied as a function of cutoff, and the connection to inverse 
scattering is demonstrated. 
\end{abstract}

\end{frontmatter}

\section{Introduction}

Perturbative expansions of the Lippmann-Schwinger equation for
the scattering amplitude using conventional
nucleon-nucleon interactions fail to converge in low partial
waves for several reasons. First is the 
strongly repulsive short-range interaction
present in all conventional potential models, which implies large contributions
from poorly constrained high-momentum components in intermediate states 
and requires
a nonperturbative treatment. 
Second is the tensor force, e.g., from pion exchange, which is singular
at short distances and requires iteration in the triplet channels.
Finally, the presence of low-energy bound states or nearly bound states in 
the S-waves imply poles in the $T$ matrix that render the perturbative
Born series divergent or nearly so.
These features greatly complicate the nuclear few- and many-body problems.

However, the nucleon-nucleon 
potential is not an observable, fixed uniquely from experiment,
and these nonperturbative features can vary widely among the infinite number of
potentials capable of accurately describing low-energy physics.
The effective field theory (EFT) philosophy leads us to exploit this freedom
by choosing inter-nucleon interactions appropriate for the physics problem
of interest.
In this regard,
a compelling alternative to conventional potentials for the
many-body problem is to use energy-independent nucleon-nucleon
interactions with a variable momentum cutoff, which are evolved
to low momentum using renormalization-group or model-space methods,  
together with consistent cutoff-dependent many-nucleon 
interactions~\cite{Vlowk1,Vlowk2,VlowkRG,Vlowkfew,Vlowknm,Bogner:2005fn}.
These low-momentum potentials, generically called ``$\vlowk$'', 
preserve the on-shell $T$ matrix
for momenta below the cutoff.\footnote{In this work, 
we use $\vlowk$ potentials with sharp cutoffs
on the relative momenta.  The use of smooth cutoffs for $\vlowk$, 
which have technical advantages for some applications, is discussed 
in Refs.~\cite{Bogner2006,Vlowksmooth}.}

Changing the momentum cutoff
shifts contributions between nuclear interactions and loop integrals
over intermediate states such that two-nucleon observables
are unchanged.
This affects the perturbativeness of an interaction and the
strength of the associated correlations in the wave functions,
which means that many-body calculations can be simplified,
particularly since Pauli blocking eliminates
at moderate densities (well below nuclear matter saturation) the impact of
the bound and nearly bound states in the S-waves.
In Ref.~\cite{Vlowknm}, it was shown that low-momentum interactions
make nuclear matter calculations perturbative in the particle-particle channel
and that the corresponding low-momentum three-nucleon interactions drive
saturation. This refutes the long-standing
conventional wisdom that the nuclear matter problem must be
nonperturbative in the inter-nucleon interactions in order to
obtain saturation.
   
In order to monitor and quantify our observations about perturbative
expansions, a method introduced by 
Weinberg~\cite{Weinberg} that focuses on eigenvalues of the 
operator $G_0(z) V$ proves very useful.
This method was applied in Ref.~\cite{Vlowknm}
at the deuteron binding energy to the S-waves
in free space and in the medium.
In the present work, we provide a complete analysis in free space.
The Weinberg eigenvalues obtained for nucleon-nucleon interactions 
are studied as a function of the cutoff
for various partial-wave channels and for a range of energies. 
When these complex eigenvalues lie within the unit circle, 
the $T$ matrix expansion in terms of the potential $V$ is perturbative,
and the largest eigenvalue determines the rate of convergence.
By focusing separately on
the ``attractive'' and ``repulsive'' eigenvalues (see
Sect.~\ref{sect:weinberg}), we can isolate the different sources of
nonperturbative behavior in nuclear forces.

In Sect.~\ref{sect:weinberg}, 
we apply this analysis to $\vlowk$ derived from
the Argonne $v_{18}$ potential~\cite{AV18}, which has been used in
the most accurate ab-initio calculations of nuclei and 
nuclear matter to date.
The removal of the strongly repulsive
core with decreasing cutoff
is tracked quantitatively via the largest repulsive Weinberg
eigenvalues.
Comparisons to other potentials (CD-Bonn~\cite{CDBonn}
and Nijmegen~\cite{Nijmegen})
shows the model dependence inherent
in the short-distance descriptions, which is removed as the cutoff
is lowered to $2\,\mbox{fm}^{-1}$.
Chiral EFT potentials~\cite{N3LO,N3LOEGM}, which provide a more 
systematic and model-independent approach to inter-nucleon 
interactions, typically have cutoffs below $3.5 \fmi$, but running 
the cutoff lower tames the impact of the tensor force and of new 
nonperturbative features entering at N$^3$LO.
The eigenvectors corresponding to the Weinberg
eigenvalues lead naturally to separable approximations to the
nucleon-nucleon interaction~\cite{BrownJackson}. 
In Sect.~\ref{sect:separable}, we show that the 
expansion in separable potentials becomes more effective for lower
cutoffs, and compare $\vlowk$ with separable inverse-scattering 
solutions. Sect.~\ref{sect:conclusions} summarizes our conclusions.

\section{Weinberg Eigenvalue Analysis for Low-Momentum Interactions}
\label{sect:weinberg}
   
The eigenvalue analysis introduced by Weinberg~\cite{Weinberg} provides
quantitative conditions for the perturbative convergence of the $T$ matrix.
This allows us to identify momentum cutoffs at which repulsive core scattering
and iterated tensor contributions become perturbative.
Here we briefly review the Weinberg formalism and refer the reader to the
literature~\cite{Weinberg,BrownJackson,Scadron,Gloeckle} for more details.

The Lippmann-Schwinger equation for the $T$ matrix with energy-independent
potential $V$ is given in operator form by
\beqn
   T(z) = V + V G_0(z) T(z) \, ,
\eeqn
where the free propagator $G_0(z)$ with kinetic energy $H_0$ is 
\beqn
   G_0(z) = \frac{1}{z - H_0} \, .
\eeqn
The $T$ matrix can be expanded in the perturbative Born series as
\beqn
   T(z) = V + VG_0(z)V + VG_0(z)VG_0(z)V + \ldots \,.
\eeqn
We are interested in the convergence of this series at different 
energies, which is closely related to the presence of bound states
or resonances.
Suppose that $V$ supports a bound state at $z = E_b < 0$. 
Since $T(z)$ has a pole at
$z = E_b$ and the individual terms in the Born series
are all finite, this implies that the series diverges as $z \rightarrow E_b$. 

We can rewrite the eigenvalue
problem for the bound state $|b \rangle$,
\beqn
   (H_0 + V) |b \rangle = E_b |b \rangle \, ,
\eeqn
as 
\beqn   
  \frac{1}{E_b - H_0} V |b \rangle = G_0(E_b) V |b \rangle
           = |b \rangle \, ,
\eeqn
which motivates 
the generalization to a new eigenvalue problem for complex energies $z$:
\beqn
  G_0(z) V \, | \Psi_{\nu}(z) \rangle = \eta_{\nu}(z) \, 
        | \Psi_{\nu}(z) \rangle \, .
   \label{eq:Weinberg}
\eeqn
Here the index $\nu$ labels the Weinberg eigenvalues
(which are always discrete~\cite{Weinberg,Gloeckle}) 
and eigenvectors. The $\eta_\nu(z)$ are analytic functions cut along
$0 \leqslant z < \infty$~\cite{Gloeckle}. For positive energies, the
eigenvalues are defined as the limits on the upper or lower edges of
the cut:
\beqn
  \lim_{\epsilon \rightarrow 0} G_0(E\pm i\epsilon) V 
  |\Psi^{(\pm)}_{\nu}(E) \rangle = \eta^{(\pm)}_{\nu}(E) \, 
  |\Psi^{(\pm)}_{\nu}(E) \rangle \, .
\eeqn
In the following, it is understood that for positive energy $\eta_\nu(E) =
\eta^{(+)}_\nu(E)$.

As demonstrated in Ref.~\cite{Weinberg}, the perturbative Born 
series for $T(z)$ diverges if and only if there is an
eigenvalue with $|\eta_{\nu}(z)| \geqslant 1$. The necessity of 
this condition is clear, since
\beqn
T(z) \, | \Psi_{\nu}(z) \rangle = \bigl( 1 + \eta_{\nu}(z)
+ \eta_{\nu}(z)^2 + \ldots \bigr) \, V \, | \Psi_{\nu}(z) \rangle
\eeqn
diverges for $|\eta_{\nu}(z)| \geqslant 1$. 
Any scattering state
$|\Psi \rangle$ can be written as a linear combination of the Weinberg
states $|\Psi_{\nu}(z) \rangle$, thus the Born series will converge 
only if all $\eta_{\nu}(z)$ lie within the unit circle. Furthermore, 
the rate of convergence is controlled by the largest $|\eta_{\nu}(z)|$, 
with smaller magnitudes implying faster convergence.
  
A rearrangement of Eq.~(\ref{eq:Weinberg}) gives a simple interpretation 
of the eigenvalues $\eta_{\nu}(z)$ in terms of the Schr\"odinger equation,
\beqn
  \bigl( H_0 + \frac{1}{\eta_{\nu}(z)} \, V \bigr) \, | \Psi_{\nu}(z) 
    \rangle = z \, | \Psi_{\nu}(z) \rangle .
\eeqn
The eigenvalue $\eta_{\nu}(z)$ can thus be viewed as an energy-dependent 
coupling that must divide $V$ to produce a solution to the 
Schr\"odinger equation at energy $z$. 
If $V$ supports a bound state at 
$z=E_B$, then there is some $\nu$ with $\eta_{\nu}(E_B)=1$, which implies 
a divergence of the Born series for nearby energies. However, what 
matters for convergence at a given energy $z$ is not simply the presence 
of nearby physical bound states, but rather the entire set of eigenstates 
that can be shifted to $z$ when the interaction is divided by 
$\eta_{\nu}(z)$.  For negative energies, a purely attractive 
$V$ gives positive $\eta_{\nu}(z)$ values, while a purely repulsive $V$ 
gives negative eigenvalues, as the sign of the interaction must be flipped
to support a bound state. For this reason, we follow convention and 
refer to negative eigenvalues as repulsive and positive ones as 
attractive. In the case of conventional nuclear interactions, the repulsive 
core generates at least one large and negative eigenvalue that causes 
the Born series to diverge in the low partial waves.

\begin{figure}[p]
  \centerline{\includegraphics*[width=3.5in]{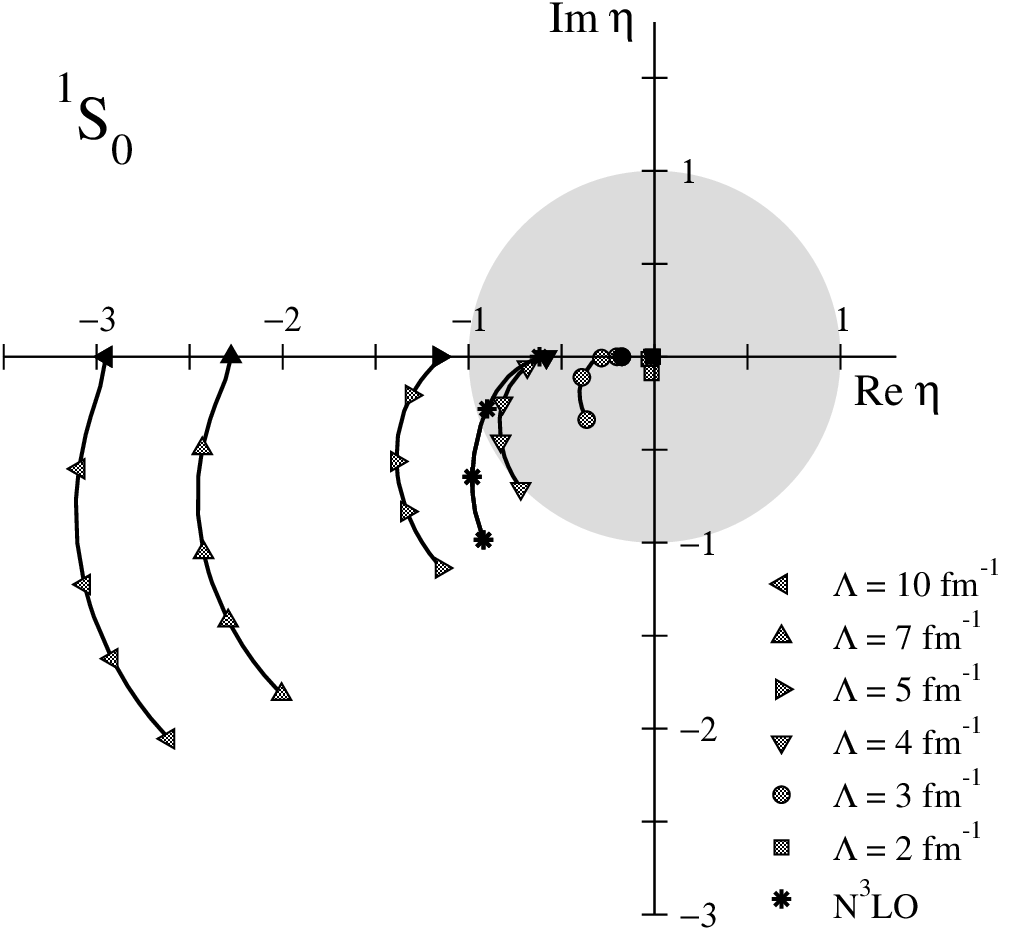}}
  \vspace*{-.1in}
  \caption{Trajectories of the largest repulsive Weinberg eigenvalue 
     in the $^1$S$_0$ channel as a function of energy 
     for $\vlowk$ derived from the Argonne $v_{18}$ potential.
     Our results for selected cutoffs are indicated by the different symbols.  
     The positions of the symbols on each trajectory mark the eigenvalues
     for center-of-mass energies $E_{\rm cm}= k^2 = 0, 25, 66, 100$ and $150 \mev$, starting
     from the filled symbol at $0 \mev$. The trajectory with stars are
     eigenvalues for the N$^3$LO potential of Ref.~\cite{N3LO}.}
  \label{fig:etaEcmRepSing}
  \vspace*{.1in} 
  \centerline{\includegraphics*[width=3.5in]{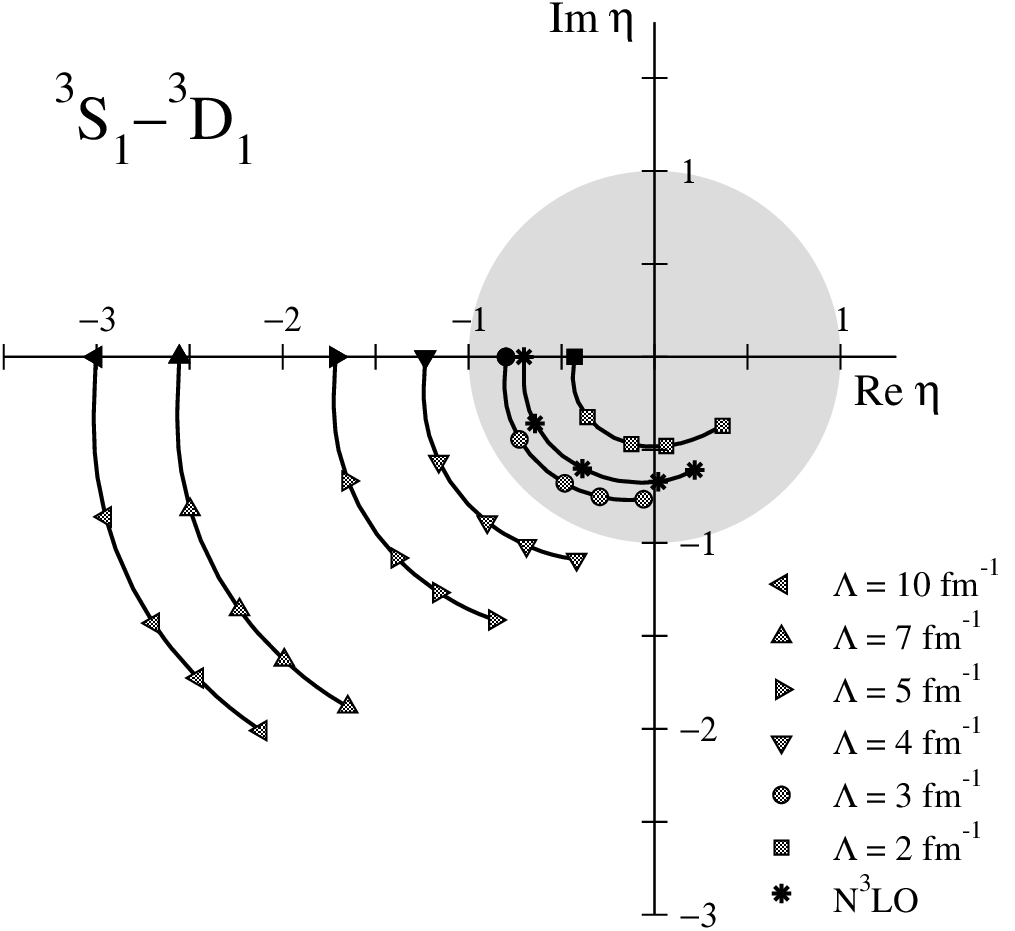}}
  \vspace*{-.1in}
  \caption{Trajectories of the largest repulsive Weinberg eigenvalue 
     in the $^3$S$_1$--$^3$D$_1$ coupled channel as a function of energy 
     for $\vlowk$ derived from the Argonne $v_{18}$ potential.
     Our results for selected cutoffs are indicated by the different symbols.     
     The positions of the symbols on each trajectory mark the eigenvalues
     for center-of-mass energies $E_{\rm cm} = 0, 25, 66, 100$ and $150 \mev$, starting
     from the filled symbol at $0 \mev$. The trajectory with stars are 
     eigenvalues for the N$^3$LO potential of Ref.~\cite{N3LO}.}
  \label{fig:etaEcmRepTrip}
\end{figure}

For energies greater than zero, the eigenvalues are still discrete
but become complex.
Weinberg also showed that there are at most a finite number of eigenvalues
with magnitudes greater than unity~\cite{Weinberg}. 
This is used in Weinberg's ``quasiparticle'' 
method to systematically isolate 
the nonperturbative parts of the potential in separable form.
The eigenvalues vary continuously with energy, so they can be plotted as
trajectories in the complex plane, as in Figs.~\ref{fig:etaEcmRepSing}
and~\ref{fig:etaEcmRepTrip}.
For each energy, the overall
largest eigenvalue determines whether or not the Born series
converges at that energy.
But by considering attractive and repulsive eigenvalues separately
we can isolate the contributions of different sources of nonperturbative
behavior.

\begin{figure}[t]
 \centerline{\includegraphics*[width=5.0in]{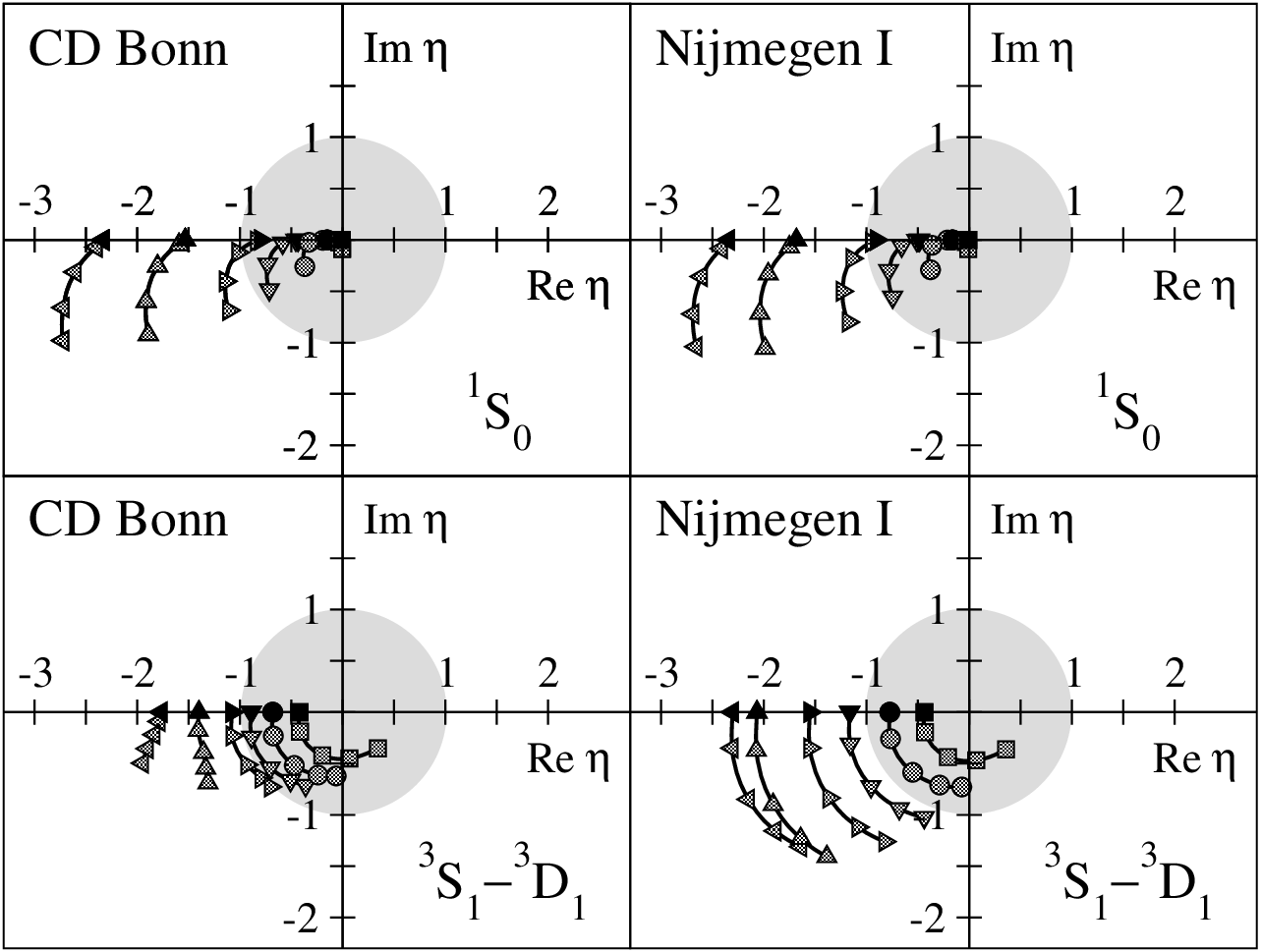}}
  \caption{Trajectories of the largest repulsive Weinberg eigenvalue 
     in the $^1$S$_0$ channel and the
     $^3$S$_1$--$^3$D$_1$ coupled channel as a function of energy 
     for $\vlowk$ derived from the CD-Bonn (left)
     and Nijmegen I potentials (right).
     Our results for selected cutoffs are indicated by the different symbols
     (see Fig.~\ref{fig:etaEcmRepSing} for the legend).     
     The positions of the symbols on each trajectory mark the eigenvalues
     for center-of-mass energies $E_{\rm cm}=0, 25, 66, 100$ and $150 \mev$, starting
     from the filled symbol at $0 \mev$.}
  \label{fig:etaEcmRepOther}
\end{figure}

Here we calculate the Weinberg eigenvalues in different partial waves 
as a function of energy for $\vlowk$ over a wide range of cutoffs.
The construction of $\vlowk$ using renormalization-group (RG) or 
model-space methods is described in detail elsewhere~\cite{Vlowk2,VlowkRG}.
In short, the RG approach starts with the partial-wave 
Lippman-Schwinger equation for the $T$
matrix with a cutoff $\Lambda$ on intermediate relative momenta. 
By imposing the condition $dT(k',k;k^2)/d\Lambda = 0$ on the
half-on-shell $T$ matrix, we ensure that low-momentum observables are independent
of the scale $\Lambda$ and generate 
energy-independent potentials. 
This leads to coupled differential equations for the interaction $\vlowk(k',k)$,
which are solved on a momentum grid suitable for Gaussian integration,
starting from 
a large initial $\Lambda_0$ appropriate for the starting ``bare'' interaction.
This starting point could be the Argonne $v_{18}$, CD-Bonn, 
or Nijmegen potential, the chiral 
EFT potentials at N$^3$LO, or any other 
potential that reproduces two-body observables.
For sharp cutoffs $\Lambda \geqslant 2 \fmi$, the resulting $\vlowk$ preserves 
the elastic phase shifts up to laboratory energies $E_{\rm lab} = 
2 \Lambda^2 \geqslant 330 \mev$ (in units where $\hbar^2/m=1$).
$\vlowk(k',k)$ is obtained on a momentum grid, where the eigenvalue problem
Eq.~(\ref{eq:Weinberg}) is given by
\beqn
  \frac{2}{\pi} \, \int\limits_0^\Lambda\,  k^2 dk \: \frac{\vlowk(k',k)}{
  p^2 - k'^2 + i \epsilon} \, \langle k | \Psi_{\nu}(p^2) \rangle =
  \eta_{\nu}(p^2) \, \langle k' | \Psi_{\nu}(p^2) \rangle \,.
  \label{eq:evproblem}
\eeqn
We solve Eq.~(\ref{eq:evproblem}) by converting it to a left-eigenvalue 
problem, where we integrate over the singularity using $1/(x+i\epsilon) 
= {\mathcal P}/x - i\pi\delta(x)$. Alternatively, we can solve the 
complex right-eigenvalue problem for $V G_0(z) \, [ V |\Psi_\nu(z)\rangle]
= \eta_\nu(z) \, [ V |\Psi_\nu(z)\rangle]$, which has the same spectrum
as $G_0(z) V$ and integrates over the singularity directly.

In Figs.~\ref{fig:etaEcmRepSing} and~\ref{fig:etaEcmRepTrip}, we show the
trajectories of the largest repulsive eigenvalue for the S-waves 
as a function of positive energy over a range of cutoffs from $\Lambda
= 10 \fmi$ to $2 \fmi$. The eigenvalues are for the $\vlowk$ evolved from
the Argonne $v_{18}$ potential, and for $\Lambda > 10 \fmi$ there are 
essentially no changes in the eigenvalues. In all cases, a trajectory 
starts on the real axis with energy $E=0$ (for repulsive eigenvalues
$\eta_\nu(0) < 0$) and evolves counter-clockwise~\cite{Weinberg}. 
If $E$ were decreased 
to negative values, the trajectory would continue along the real axis 
with decreasing magnitude of $\eta$ as $E$ becomes more negative. 
It is evident that the magnitude of the largest repulsive eigenvalue at
all energies is greatly cutoff dependent.  
The decrease with lowering the cutoff reflects the elimination
of the repulsive core of the potential.  
In the singlet channel, Fig.~\ref{fig:etaEcmRepSing},
the trajectory lies completely inside the shaded
unit circle for cutoffs near $4 \fmi$ and below,
which implies that the Born series becomes perturbative with respect
to the repulsive core (but still converges very slowly at low energies
due to the nearly bound state at threshold, see the discussion on
the attractive Weinberg eigenvalue and Fig.~\ref{fig:etaEcmAttSing} below).
By $\Lambda = 2 \fmi$ the largest repulsive eigenvalue is very small at all 
energies, which shows that the nonperturbative behavior from the repulsive 
core has been eliminated and has little impact on the convergence.

In the triplet channel, Fig.~\ref{fig:etaEcmRepTrip}, where the tensor 
contribution is active, 
the repulsive core is still the dominant source for the largest
repulsive eigenvalue.
When the cutoff is lowered to eliminate the core, the repulsive tensor
interaction is left as the now-dominant feature in the triplet channel,
so that the largest repulsive eigenvalues are just within the unit circle
for $\Lambda = 3 \fmi$ and still have a magnitude of $\eta \approx 0.5$ for
$\Lambda = 2 \fmi$.
However, the decrease from $\Lambda = 3 \fmi$ to $2 \fmi$
is significant for ensuring a convergence of
particle-particle ladders in the nuclear medium, since second-order 
tensor contributions are known to strongly excite intermediate-state 
momenta peaked at $k \approx 2.5-3.5 \fmi$ in nuclear matter~\cite{GerryMBbook}. 
In Fig.~\ref{fig:etaEcmRepOther} we show
the largest repulsive Weinberg eigenvalues of $\vlowk$ obtained from
the CD-Bonn and Nijmegen I potentials. The eigenvalues for these
energies are nearly identical for cutoffs below $\Lambda = 3 \fmi$, 
whereas the coalescence of all momentum-space matrix elements of 
$\vlowk$ occurs for lower cutoffs $\Lambda \sim 2 \fmi$~\cite{Vlowk1,Vlowk2}.
We further observe that the singlet channel eigenvalues are very
similar even for larger cutoffs, but the triplet eigenvalues
differ substantially for large cutoff. 

\begin{figure}[p]
 \centerline{\includegraphics*[width=2.8in]{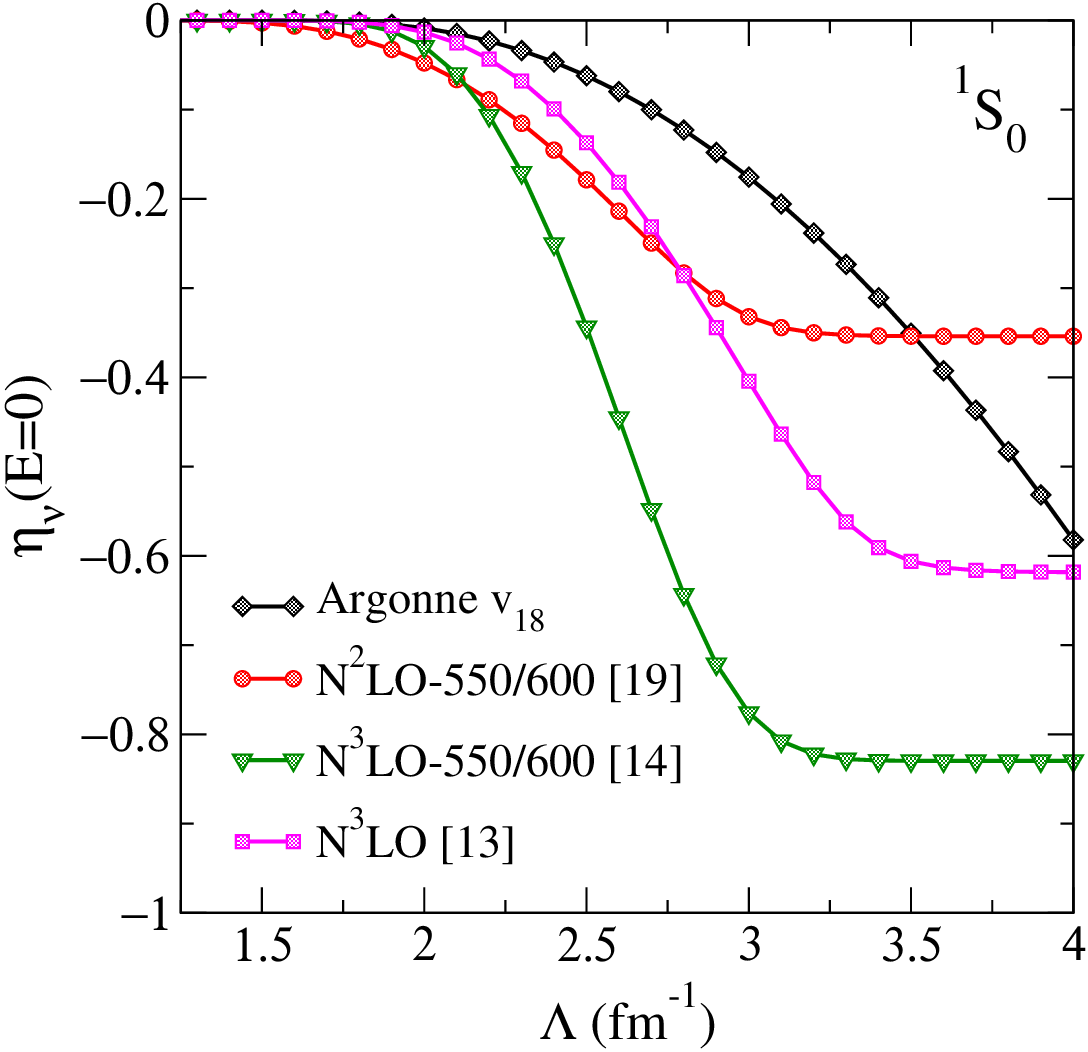}
             \includegraphics*[width=2.6in]{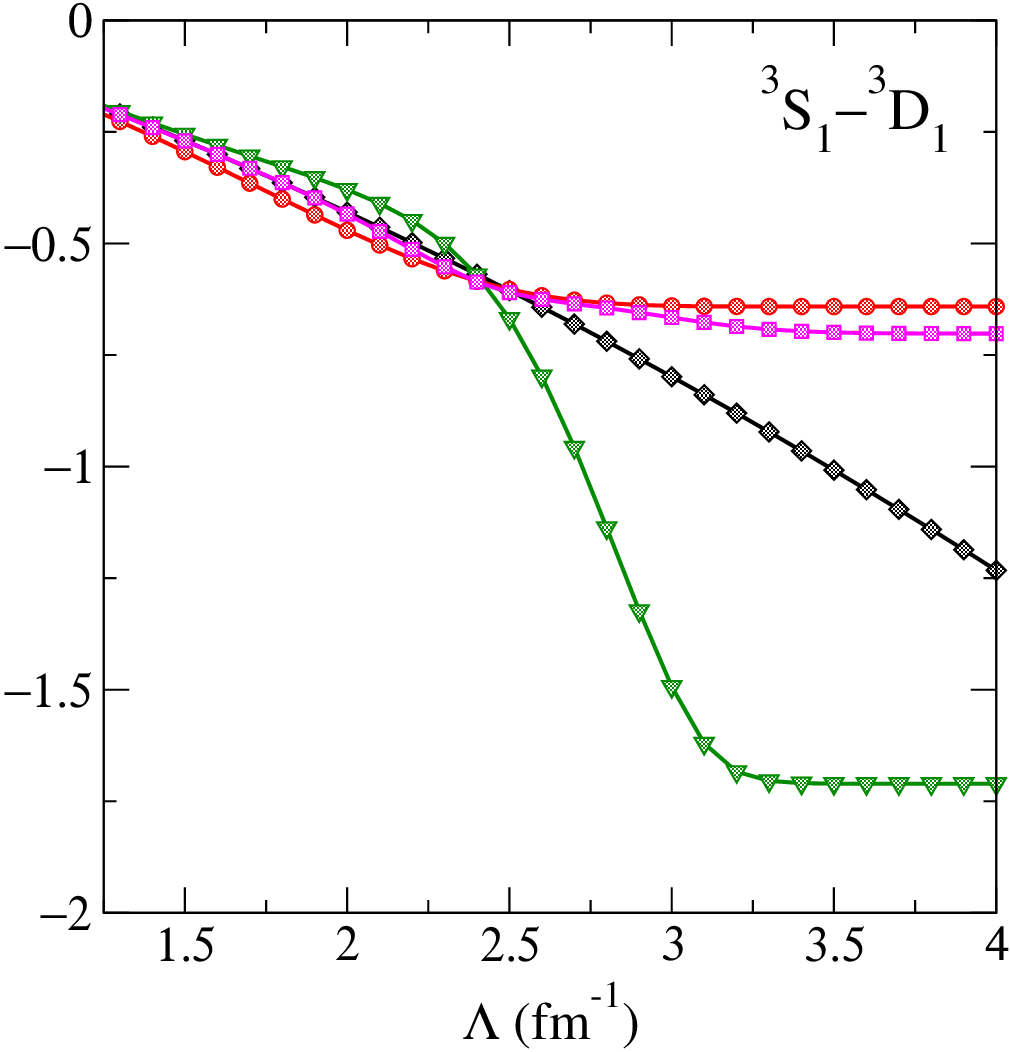}}
  \caption{The largest repulsive Weinberg eigenvalues for $E=0$ in the 
     $^1$S$_0$ channel (left) and the $^3$S$_1$--$^3$D$_1$ coupled channel 
     (right) as 
     a function of cutoff for $\vlowk$ derived from chiral interactions.
     Results are shown for the N$^3$LO potential of Entem and
     Machleidt~\cite{N3LO}, for the N$^3$LO potential of Epelbaum
     \emph{et al.}~\cite{N3LOEGM} with different cutoffs 
     $\Lambda$/$\tilde{\Lambda}$ (as indicated in MeV), 
     and for the N$^2$LO potential~\cite{N2LO}. 
     For comparison, we have plotted the largest
     repulsive Weinberg eigenvalues for $\vlowk$ derived from the
     Argonne $v_{18}$ potential.}
  \label{fig:chiral}
\end{figure}

\begin{figure}[p]
 \centerline{\includegraphics*[width=2.8in]{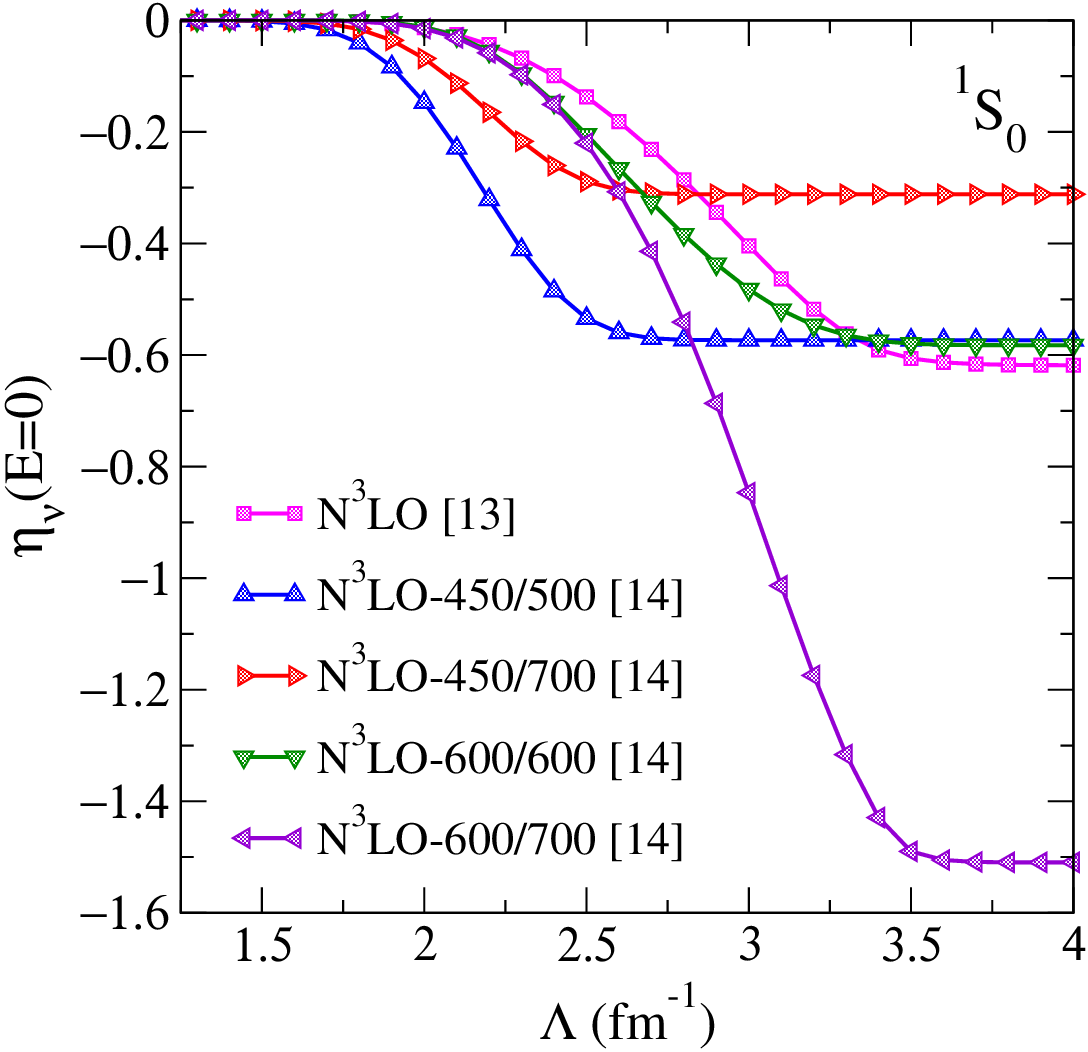}
             \includegraphics*[width=2.6in]{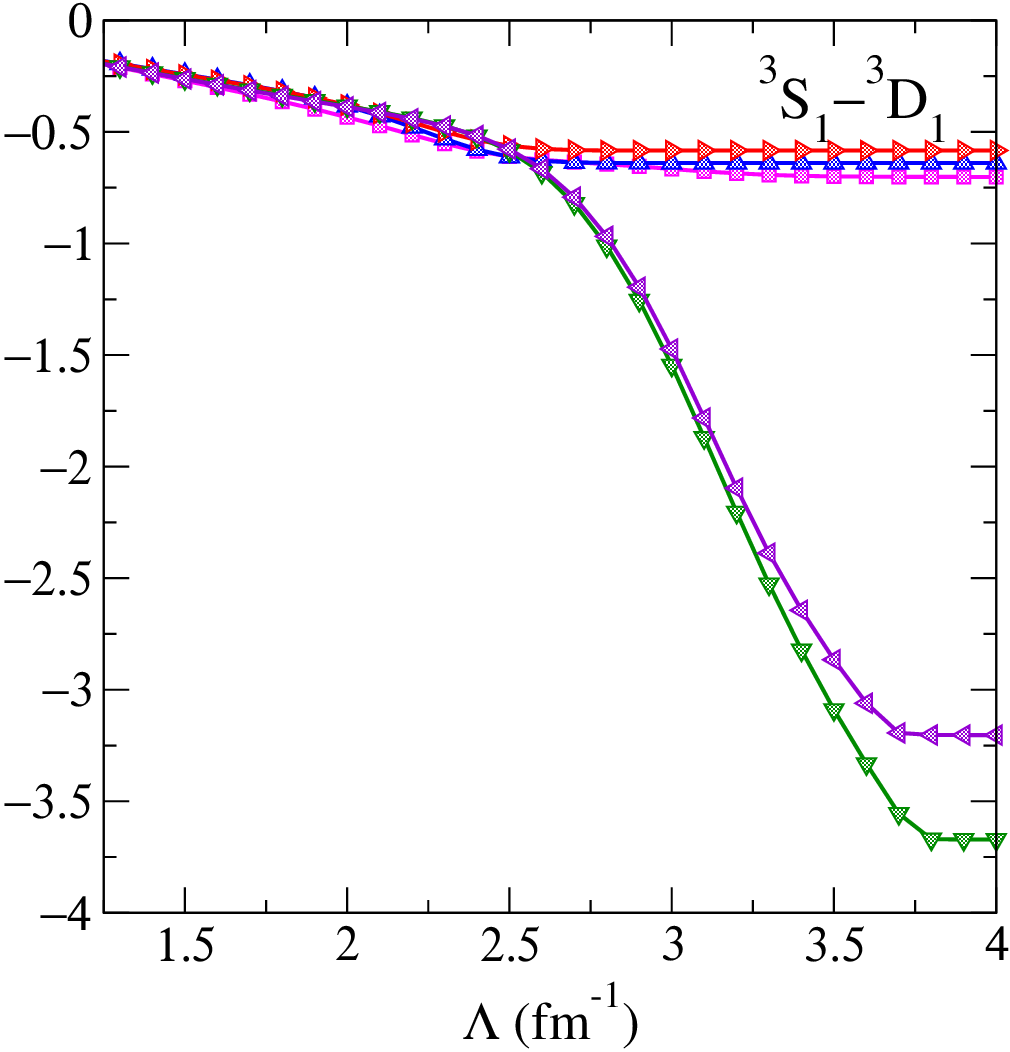}}
  \caption{The largest repulsive Weinberg eigenvalues for $E=0$ in the 
     $^1$S$_0$ channel (left) and the $^3$S$_1$--$^3$D$_1$ coupled channel 
     (right) as 
     a function of cutoff for $\vlowk$ derived from chiral interactions.
     Results are shown for the N$^3$LO potential of Entem and
     Machleidt~\cite{N3LO} and for the N$^3$LO potential of Epelbaum
     \emph{et al.}~\cite{N3LOEGM} with different cutoffs 
     $\Lambda$/$\tilde{\Lambda}$ (as indicated in MeV).}
  \label{fig:chiral2}
\end{figure}

Chiral EFT interactions typically have cutoffs below $3.5 \fmi$ and 
therefore are expected to be soft potentials. However, as shown in
Fig.~\ref{fig:etaEcmRepSing}, the N$^3$LO potential of Entem and
Machleidt~\cite{N3LO} has substantial repulsive eigenvalues in the 
singlet channel, which are close to the $\vlowk$ results for $\Lambda = 
4 \fmi$. Therefore, we study the behavior of chiral EFT interactions
in more detail in Figs.~\ref{fig:chiral} and \ref{fig:chiral2}, 
using Weinberg eigenvalues
as a diagnostic. Our results show that
the repulsive eigenvalues are large for the N$^3$LO potential of 
Epelbaum \emph{et al.}~\cite{N3LOEGM}, and that there is a major
decrease as the cutoff is run down from $\Lambda = 3 \fmi$ to 
$2 \fmi$. This decrease is important, as the magnitude of the 
largest repulsive eigenvalue increases in the $^1$S$_0$ channel
for $E > 0$ (see Fig.~\ref{fig:etaEcmRepSing}) and because the 
tensor force in the $^3$S$_1$--$^3$D$_1$ coupled channel is
softened when chiral interactions are evolved to lower cutoffs
(see also Fig.~\ref{fig:etaEcmRepTrip}). 

In addition to the tensor contribution at all orders, there 
are new nonperturbative sources due to the central parts of
pion exchanges that enter at N$^3$LO. This can be seen from
the S-wave component of the deuteron wave function at N$^3$LO, 
Fig.~13 in Ref.~\cite{N3LOEGM}, where the very short-range part 
for $\Lambda = 550 \mev$ is similar to what is found with a
repulsive core (even for $\Lambda = 450 \mev$, the S-wave deuteron 
wave function lies between the bare Argonne $v_{18}$ potential
and $\vlowk$ for $\Lambda = 4 \fmi$). This observation and the 
comparison in Fig.~\ref{fig:chiral} to the eigenvalues of the 
N$^2$LO potential~\cite{N2LO} suggest that the large repulsive Weinberg 
eigenvalues at N$^3$LO are due to nonperturbative features in the 
central part of the sub-sub-leading $2\pi$-exchange interaction. This
might be understood from the short-distance behavior of the 
isoscalar central
$2\pi$-exchange two-loop contributions, discussed in~\cite{Kaiser}. 
The repulsive eigenvalues 
of the N$^3$LO potential of Entem and Machleidt~\cite{N3LO},
which uses $\Lambda = 500 \mev$,
are smaller and close to the $\Lambda = 450 \mev$ 
N$^3$LO potentials of Epelbaum \emph{et al.}~\cite{N3LOEGM}. 
The differences in implementation between 
\cite{N3LO} and \cite{N3LOEGM}, such as the fitting of low-energy
constants and the regularization of two-pion 
exchange at N$^3$LO are reflected in the eigenvalues.
As seen in Fig.~\ref{fig:chiral2}, however,
the patterns of eigenvalues are not completely
systematic and require further study, which 
is deferred to a future publication. 
In any case, Figs.~\ref{fig:chiral} and \ref{fig:chiral2} 
clearly show that it is advantageous to
evolve chiral EFT interactions to lower cutoffs using the RG,
since the singular central (at N$^3$LO) and the singular tensor 
parts are softened for lower cutoffs.

\begin{figure}[p]
  \centerline{\includegraphics*[width=5.25in]{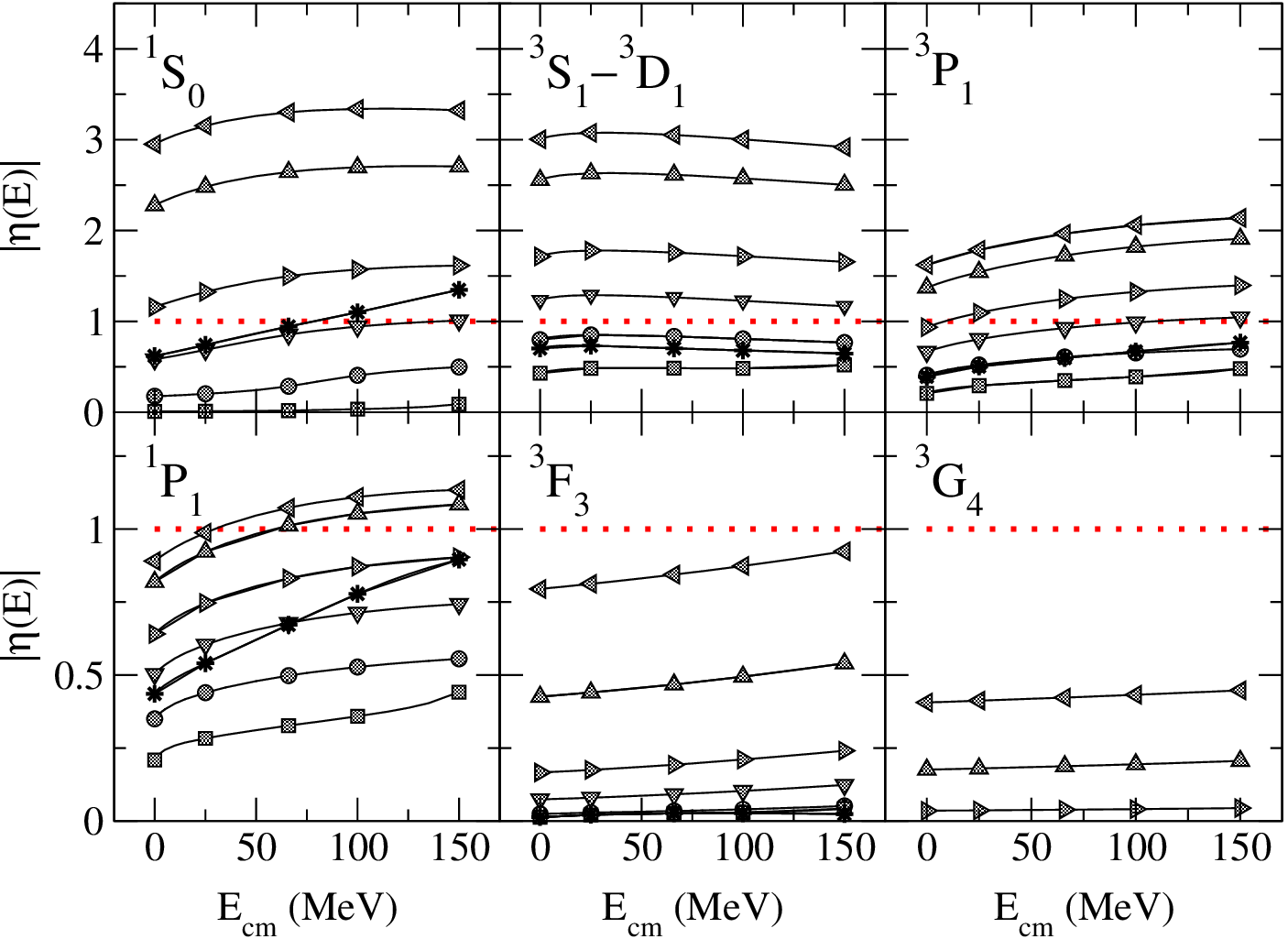}}
  \vspace*{-.1in}
  \caption{The magnitude of the
     largest repulsive Weinberg eigenvalue as a function of 
     center-of-mass energy $E_{\rm cm}$ 
     in selected channels for a range of cutoffs.  The symbols label the
     same cutoffs and the N$^3$LO potential~\cite{N3LO} 
     as in Fig.~\ref{fig:etaEcmRepSing}.}
  \label{fig:etaEcmRep}
  \vspace*{.1in} 
  \centerline{\includegraphics*[width=5.25in]{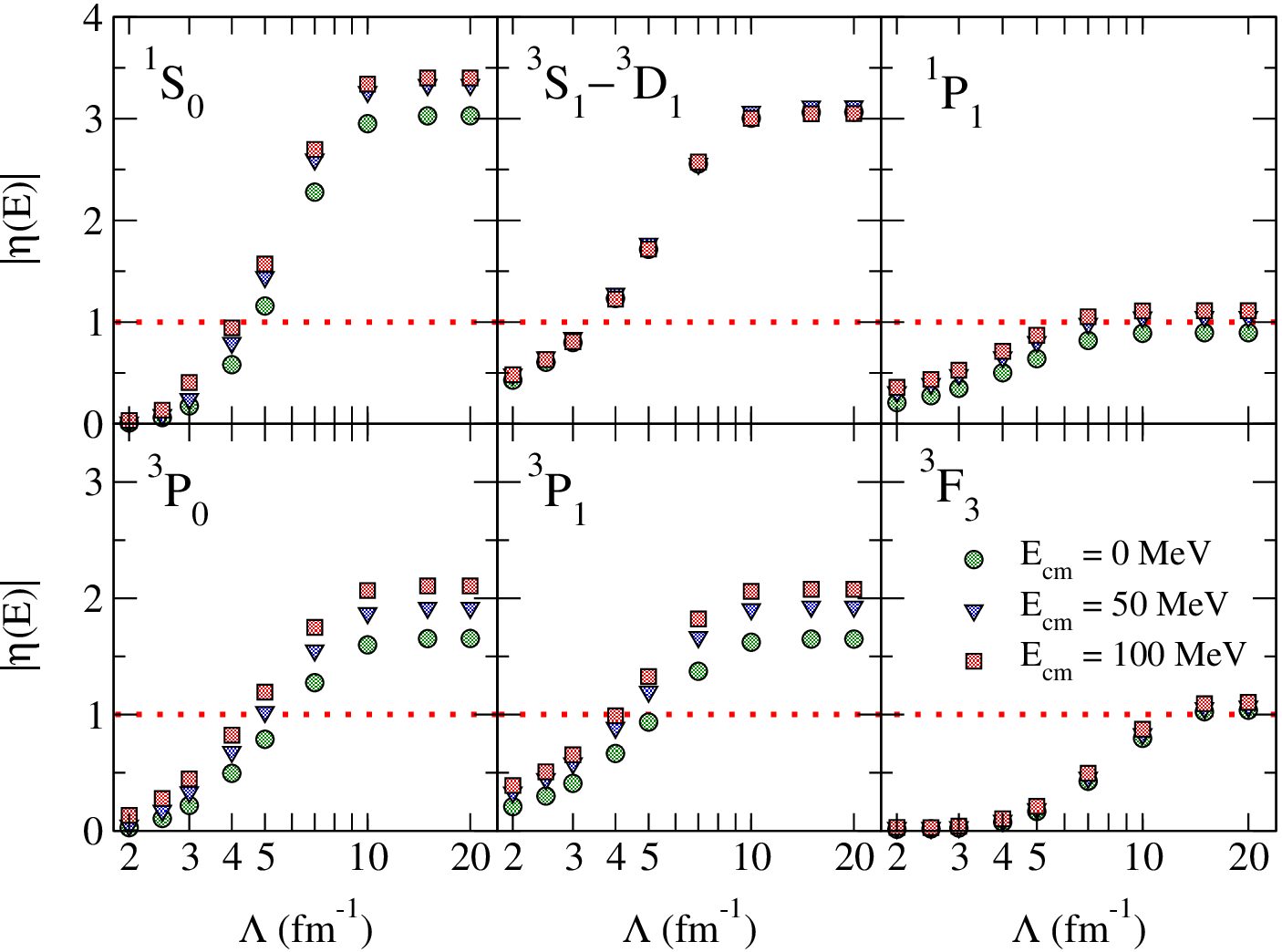}}
  \vspace*{-.1in}
  \caption{The magnitude of the
     largest repulsive Weinberg eigenvalue as a function of 
     cutoff $\Lambda$ 
     in selected channels for three center-of-mass energies.}
  \label{fig:etaLambdaRep}
\end{figure}

\begin{figure}[p]
  \centerline{\includegraphics*[width=3.5in]{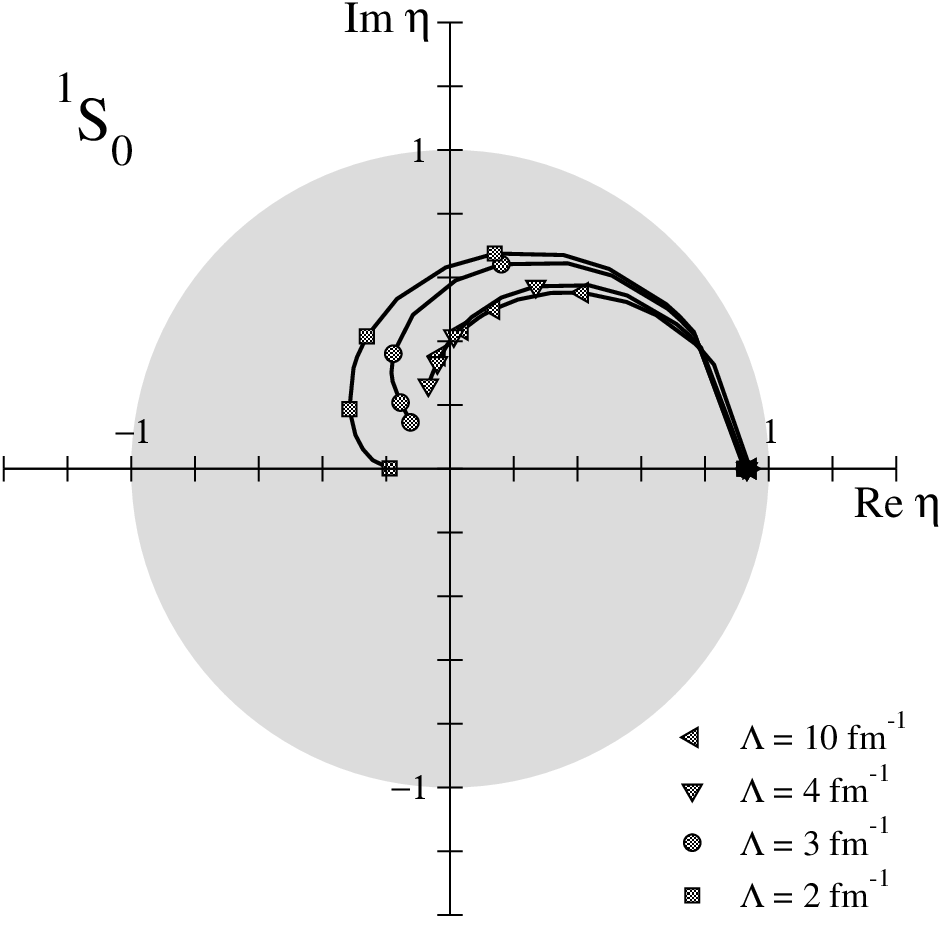}}
  \vspace*{-.1in}
  \caption{Trajectories of the largest attractive Weinberg eigenvalue 
     in the $^1$S$_0$ channel as a function of energy 
     for $\vlowk$ derived from the Argonne $v_{18}$ potential.
     Our results for selected cutoffs are indicated by the different symbols.     
     The positions of the symbols on each trajectory mark the eigenvalues
     for center-of-mass energies $E_{\rm cm}=0, 25, 66, 100$ and $150 \mev$, starting
     from the filled symbol at $0 \mev$.}
  \label{fig:etaEcmAttSing}
  \vspace*{.1in}
  \centerline{\includegraphics*[width=3.5in]{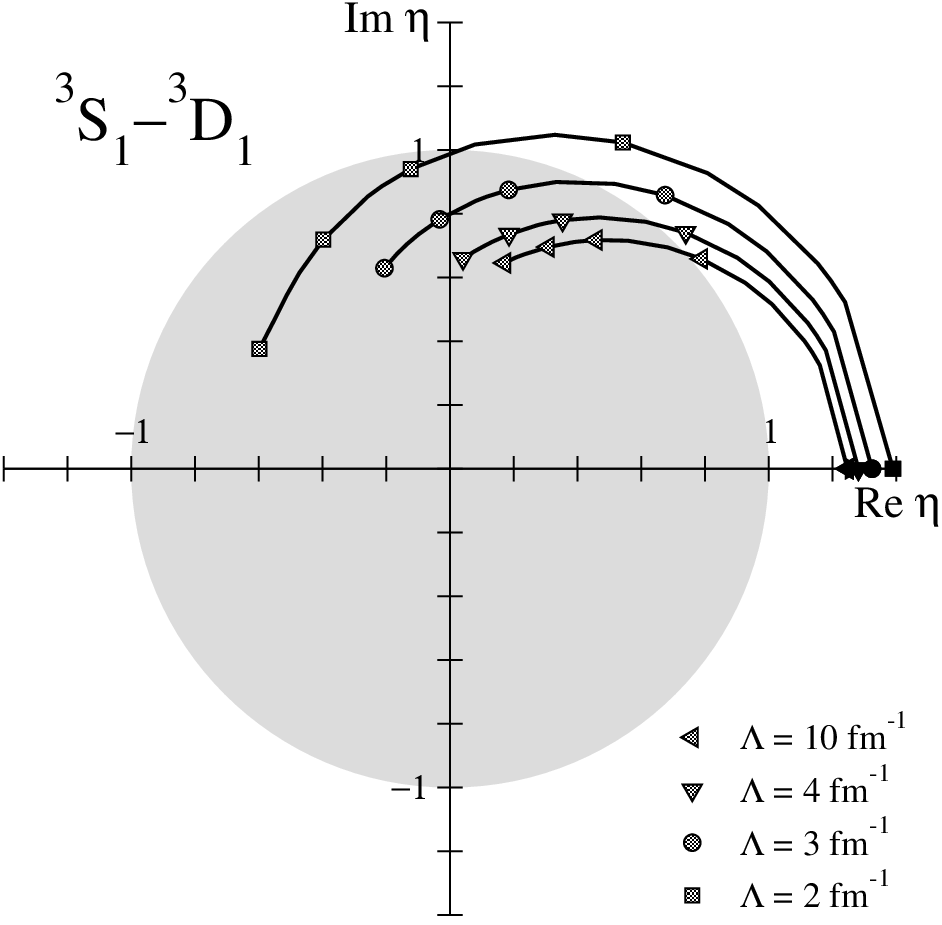}}
  \vspace*{-.1in}
  \caption{Trajectories of the largest attractive Weinberg eigenvalue 
     in the $^3$S$_1$--$^3$D$_1$ coupled channel as a function of energy 
     for $\vlowk$ derived from the Argonne $v_{18}$ potential.
     Our results for selected cutoffs are indicated by the different symbols.     
     The positions of the symbols on each trajectory mark the eigenvalues
     for center-of-mass energies $E_{\rm cm}=0, 25, 66, 100$ and $150 \mev$, starting
     from the filled symbol at $0 \mev$.}
  \label{fig:etaEcmAttTrip}
\end{figure}

\begin{figure}[t]
  \centerline{\includegraphics*[width=5.5in]{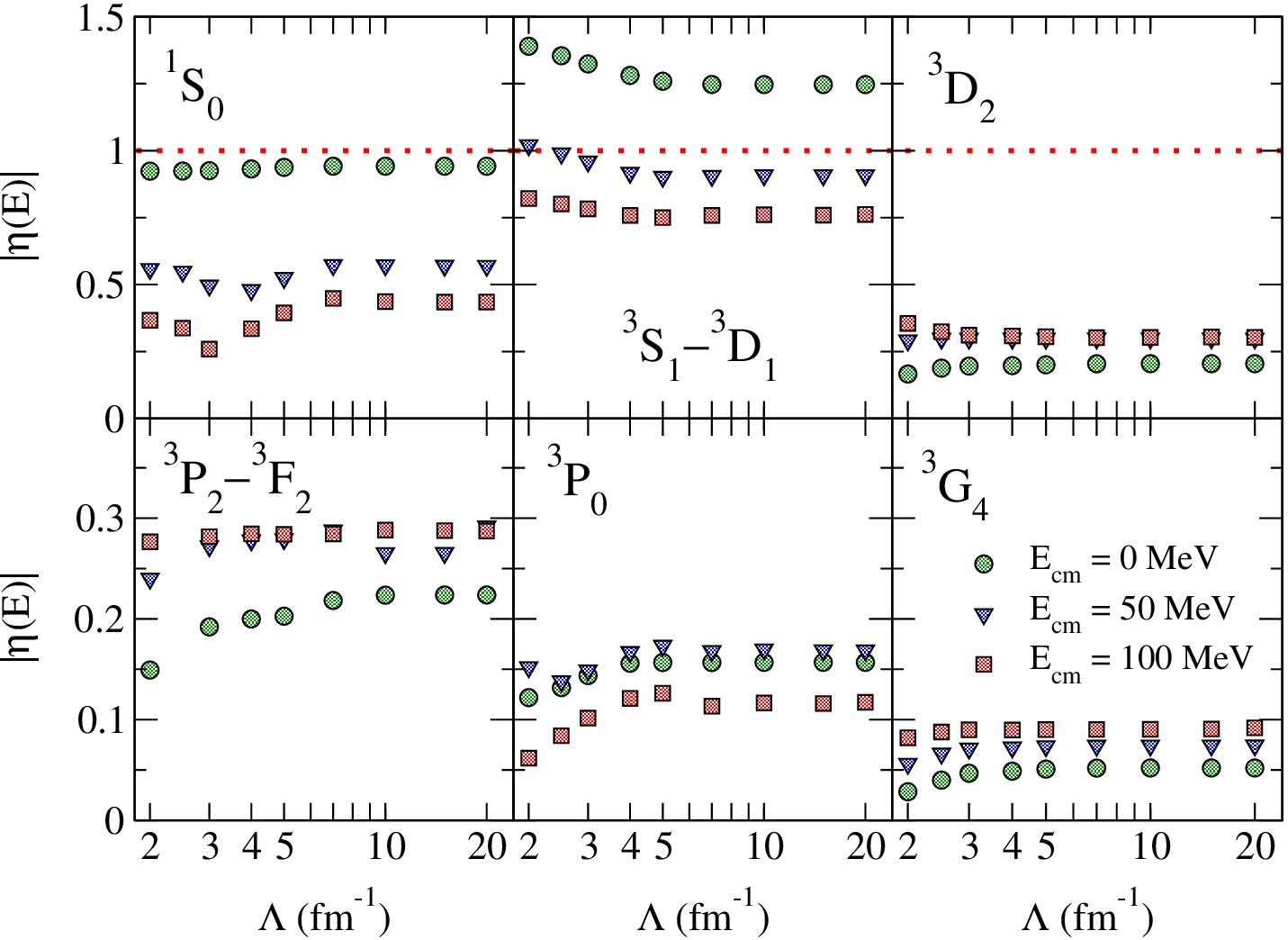}}
  \vspace*{-.1in}
  \caption{The magnitude of the
     largest attractive Weinberg eigenvalue as a function of $\Lambda$ 
     in selected channels for three laboratory energies.
     There are no significant attractive Weinberg eigenvalues for any
     $\Lambda$ in the $^1$P$_1$, $^3$P$_1$ and $^3$F$_3$ channels and
     where not plotted, the eigenvalue was too small in magnitude to 
     unambiguously identify.}
  \label{fig:etaLambdaAtt}
\end{figure}

In Figs.~\ref{fig:etaEcmRep} and~\ref{fig:etaLambdaRep}, 
the magnitudes of the largest repulsive
eigenvalues for the Argonne $v_{18}$ potential
are shown for a set of representative partial waves. The results
are plotted as a function of energy for selected cutoffs and 
as a function of cutoff for selected energies, respectively 
(where not plotted, the eigenvalue was too small in magnitude to 
unambiguously identify).
The same monotonic decrease in magnitude with decreasing cutoff
is seen in every channel.  The crossover from the nonperturbative ($|\eta| > 1$)
to the perturbative regime ($|\eta| < 1$) happens at different cutoffs in different
channels, but always for $\Lambda > 3\fmi$.
For completeness,
we note that the second-largest repulsive eigenvalue is below unit 
magnitude in all channels for all cutoffs, but shows the same rapid decrease 
to very small values by $\Lambda = 2\fmi$.

Next, we study the largest attractive Weinberg eigenvalues. Their trajectories
for the S-waves are plotted as a function of energy in 
Figs.~\ref{fig:etaEcmAttSing} and~\ref{fig:etaEcmAttTrip}.
The values $\eta(E=0) \approx 1$ reflect the presence of a 
bound state close to zero energy; if the potential were scaled in each
case to $V/\eta(0)$, there would be a bound state exactly at zero energy.
Conversely, all nuclear interactions have a unit eigenvalue $\eta(E_{\rm d})=1$ 
in the triplet channel, corresponding to the deuteron binding energy
$E_{\rm d} = -2.225 \mev$.
We find that the attractive eigenvalues increase insubstantially in magnitude 
as $\Lambda$ decreases. This small increase is similar for all energies.
As expected, the eigenvalue close to unit magnitude persists as
we lower the cutoff. This is because it is dictated by physics, namely the
near-zero-energy bound states.
However, we have shown in Ref.~\cite{Vlowknm} that the large attractive
eigenvalues decrease rapidly as a function of density due to Pauli blocking.
In contrast, the large repulsive eigenvalues for large cutoffs are
little affected by Pauli blocking.

The role of the tensor force can be seen by comparing the $^3$P$_0$ and
$^3$P$_1$ channels in Figs.~\ref{fig:etaLambdaRep} and \ref{fig:etaLambdaAtt}.
The comparison shows similar repulsive
eigenvalues (in magnitude) at large cutoff, where the dominant effect is
the short-range repulsion.  At the smallest cutoffs, $^3$P$_0$ has tiny
repulsive eigenvalues while the $^3$P$_1$ repulsive eigenvalues are
non-negligible.  This is consistent with tensor contributions
being attractive in $^3$P$_0$ and repulsive in $^3$P$_1$. 
Figure~\ref{fig:etaLambdaAtt} shows that the largest attractive eigenvalues
for channels besides the S-waves are where the long-range one-pion
exchange tensor force is attractive 
($^3$P$_0$, $^3$P$_2$--$^3$F$_2$, $^3$D$_2$ and $^3$G$_4$).  
The eigenvalues in the channels not shown
are very small in magnitude. The behavior of the
attractive eigenvalues is similar for the other potential models.

\section{Separable Approximations to the Nucleon-Nucleon Interaction}
\label{sect:separable}

The Weinberg eigenvalue analysis leads naturally to separable
approximations for nucleon-nucleon
potentials, which may be of practical interest for solving few-body
problems such as four-body scattering~\cite{Fonseca,CSB}. The 
Weinberg eigenstates $\{ | \Psi_\nu \rangle \}$ can be used 
as a basis for a separable representation of the
potential~\cite{Weinberg,BrownJackson,RAWITSCHER91,CANTON91},
with various possible ways to construct such a representation.
Here we do not attempt to optimize such an expansion, 
but focus on the efficacy of a \emph{given} separable expansion
as a function of the cutoff, and choose for illustration the
expansion in Weinberg eigenvectors at zero energy.

For $E \leqslant 0$, we use the following separable representation
\beqn
V = \sum_{\nu=1}^\infty V \, |\Psi_\nu(E) \rangle \langle \Psi_\nu(E)| 
\, V \,,
\label{eq:Vseparable}
\eeqn
where we choose the normalization of the Weinberg states such that
\beqn
\langle \Psi_{\nu'}(E) |V| \Psi_\nu(E) \rangle = \delta_{\nu'\nu} \,.
\eeqn
This separable representation is easily verified by substituting 
it into the eigenvalue problem, Eq.~(\ref{eq:Weinberg}).
The $T$ matrix is then given by
\beqn
T = \sum_{\nu=1}^\infty \frac{V|\Psi_\nu(E) \rangle \langle \Psi_\nu(E)| \, V}
{1 - \eta_{\nu}(E)} \,.
\label{eq:Tseparable}
\eeqn
As in Refs.~\cite{Weinberg,BrownJackson}, 
Eqs.~(\ref{eq:Weinberg}) and (\ref{eq:Tseparable}) lead to a sum rule for
the exact $T$ matrix at energy $E$,
\beqn
{\rm Tr\,} \bigl[ G_0(E) \, T(E) \, G_0(E) \, T(E) \bigr] =
\sum_{\nu=1}^\infty \frac{1}{\bigl(1-1/\eta_\nu(E)\bigr)^2} \,.
\label{eq:sumrule} 
\eeqn
The summation in the corresponding sum rule to a 
rank-$n$ approximation $T^{(n)}$ runs only from $\nu=1$ to $\nu=n$. 
For $E \leqslant 0$ each term in the sum is
positive definite, so the sum rule is most saturated and
the difference between $T^{(n)}$ and $T$ minimized by
choosing the $n$ eigenvalues in order of ascending
$\bigl(1-1/\eta_\nu(E)\bigr)^2$.  
This illustrates a difference between the convergence of
the separable expansion and that of the Born series (which depends on the
eigenvalues of largest magnitude rather than those closest to unity).

\begin{figure}[p]
  \centerline{\includegraphics*[width=5in]{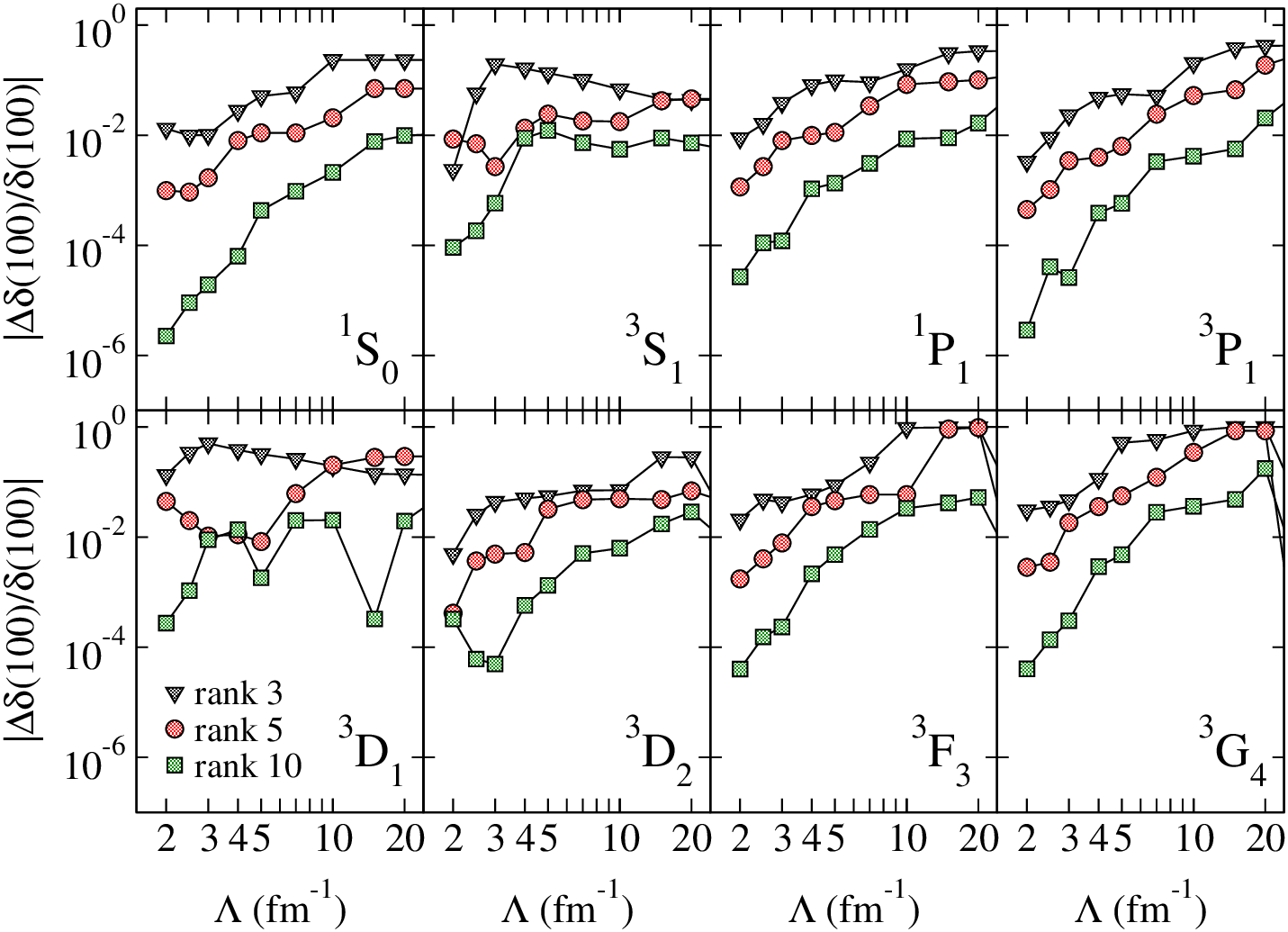}}
  \vspace*{-.1in}
  \caption{Relative error in the phase shifts for $E_{\rm lab} = 100 \mev$
   as a function of $\Lambda$ based on separable potentials generated from the
   Weinberg eigenvectors at $E=0$ corresponding to the largest $n=3, 5$ and
   $10$ eigenvalues.}
  \label{fig:pserrLambdaOne}
  \vspace*{.1in} 
  \centerline{\includegraphics*[width=5in]{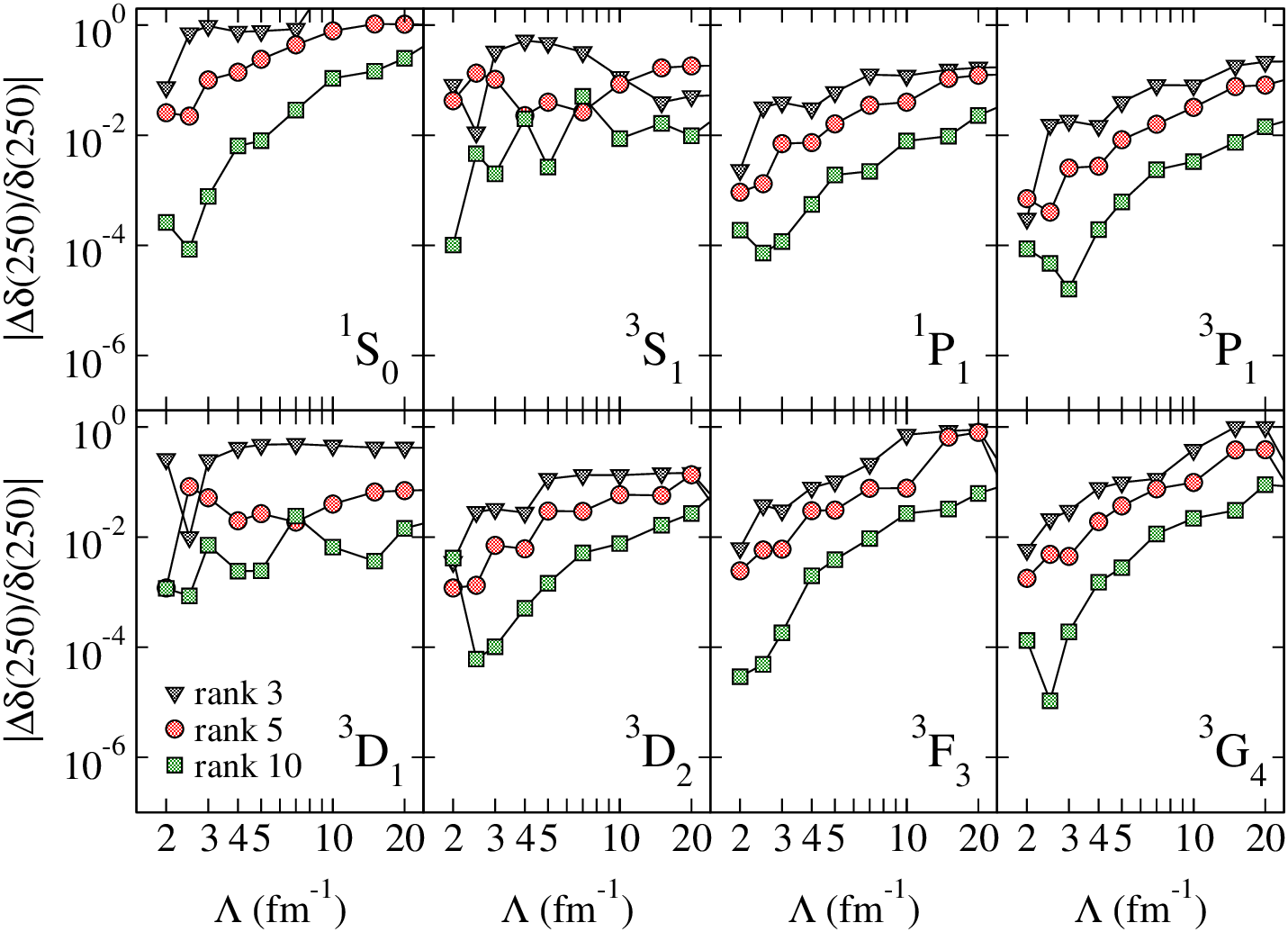}}
  \vspace*{-.1in}
  \caption{Relative error in the phase shifts for $E_{\rm lab} = 250 \mev$
   as a function of $\Lambda$ based on separable potentials generated from the
   Weinberg eigenvectors at $E=0$ corresponding to the largest $n=3, 5$ and
   $10$ eigenvalues.}
  \label{fig:pserrLambdaTwo}
\end{figure}

For practical calculations, we can apply the rank-$n$ approximation
at fixed energy $E$ to calculate the $T$ matrix at all energies $E'$.
In this case, the analysis of convergence is no longer simply given 
by the eigenvalues, but we expect the general trends to carry over.
Here we choose $E=0$ and study the accuracy of observables calculated 
from the eigenvectors corresponding to the $n$ largest eigenvalues,
using the representation Eq.~(\ref{eq:Vseparable}). As the rank $n$ 
increases, all eigenvalues that are not included decrease, and thus 
our expansion improves. In addition, we examine how the rank-$n$ 
approximation performs as a function of the cutoff, where $\vlowk$
is evolved from the Argonne $v_{18}$ potential for all results
presented in this section.

\begin{figure}[pt]
  \centerline{\includegraphics*[width=5.0in]{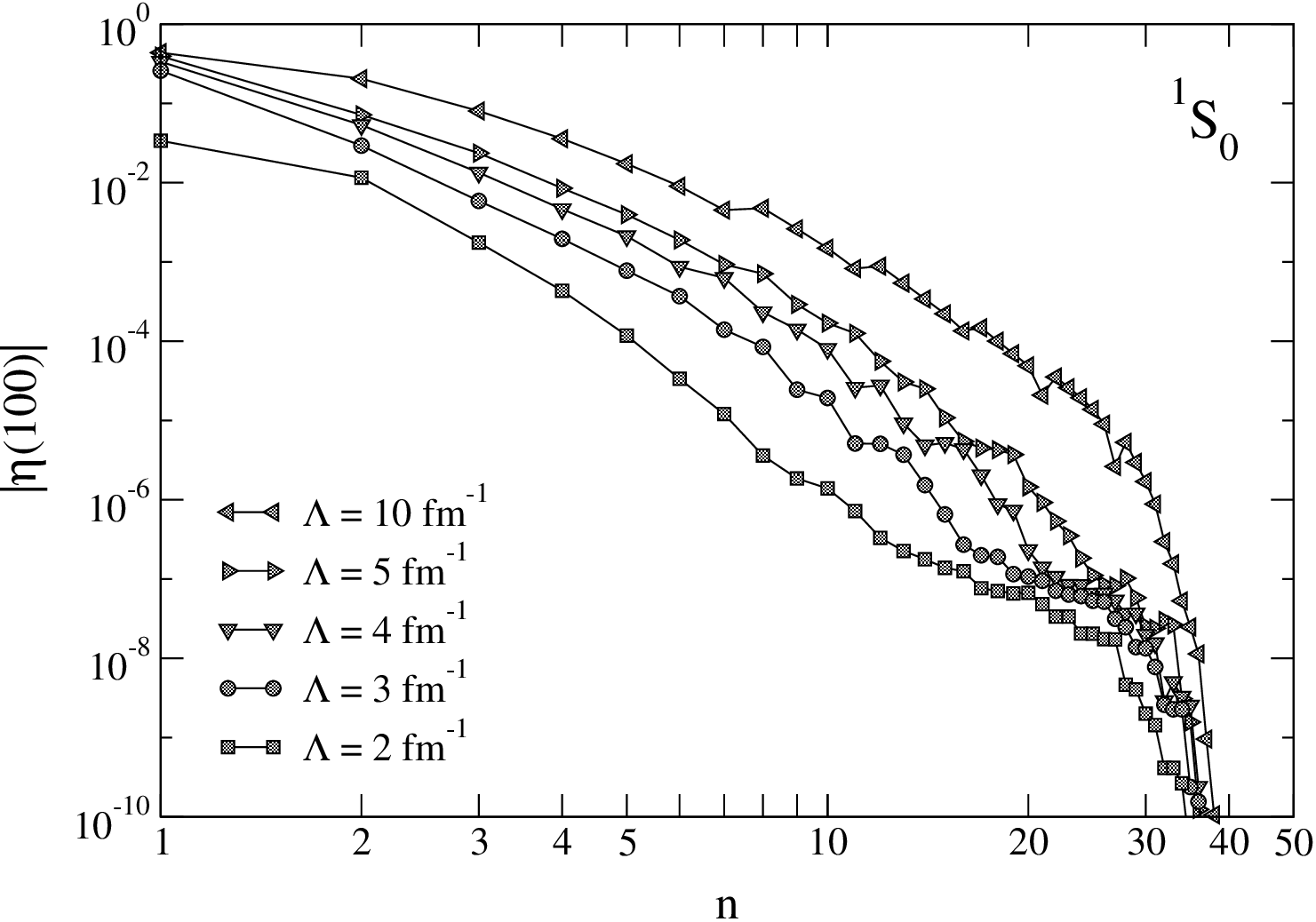}}
  \vspace*{-.1in}
  \caption{Magnitude of Weinberg eigenvalues in descending order
  for the $^1$S$_0$ channel at $E_{\rm lab} = 100 \mev$.}
  \label{fig:eta_vs_n_1s0_100}
  \vspace*{.1in} 
  \centerline{\includegraphics*[width=5.0in]{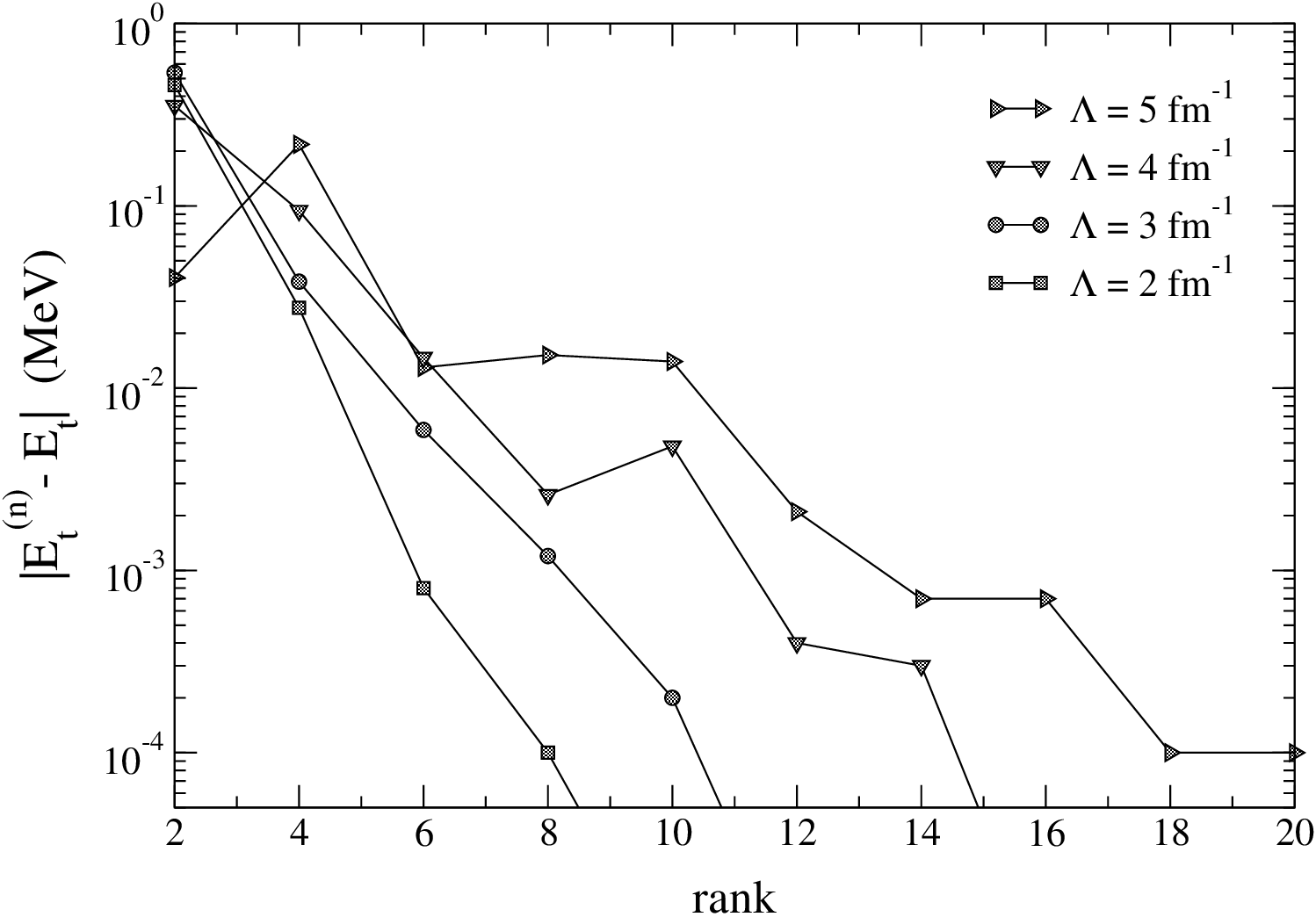}}
  \vspace*{-.1in}
  \caption{Absolute error of the triton binding energy for
  separable expansions of different rank. The results for
  different cutoffs are obtained in a fixed basis of harmonic
  oscillators (based only on two-body interactions).
  The error is with respect to the exact result for the
  same $\Lambda$ and model space $N_{\rm max}$.}
  \label{fig:Etvsrank}
\end{figure}

In Figs.~\ref{fig:pserrLambdaOne} and~\ref{fig:pserrLambdaTwo}, the
convergence of our truncation is shown by the relative error
in the calculated phase shifts for $E_{\rm lab} = 100 \mev$ and 
$250 \mev$ (which typify the full range of energies).    
Our results demonstrate that using a rank-$n$ separable approximation 
at the particular energy $E = 0 \mev$ does reproduce phase shifts 
reasonably well for $E' \neq E$. Except in a few cases, the convergence 
improves monotonically as more terms are included in the separable 
approximation and, for a particular rank-$n$ approximation, lower
cutoffs yield better convergence. As expected, we observe that the
convergence is slower in low partial waves where the tensor force 
is active ($^3$S$_1$ and $^3$D$_1$ in Figs.~\ref{fig:pserrLambdaOne} 
and~\ref{fig:pserrLambdaTwo}).
Equation~(\ref{eq:sumrule}) implies that the fall-off of Weinberg
eigenvalues should be tied to the rate of convergence.
This is consistent with Fig.~\ref{fig:eta_vs_n_1s0_100}, which shows this
fall-off in the $^1$S$_0$ channel for $E_{\rm lab} = 100 \mev$.

Further insight into how the convergence varies with the cutoff 
comes from the predicted triton binding energy.
Since we are interested in convergence rather than absolute predictions,
we consider for simplicity and convenience the expectation value of the
Hamiltonian at each $\Lambda$ in a basis of harmonic oscillators
up to $N_{\rm max} = 40$ with fixed oscillator parameter $b = 1.7 \,
\text{fm}$ (that is, the energy is not minimized with respect to 
this parameter).
The result for the triton binding energy $E_t$
with the full potential is compared to results with different
rank separable approximations at the same cutoff.
The absolute errors are plotted in Fig.~\ref{fig:Etvsrank} as a
function of rank for a variety of cutoffs.
(Note that full predictions for the triton will vary with the cutoff,
because we do not include the $\Lambda$-dependent consistent three-body
force, but for our purposes this is not relevant.)

At very low rank, the errors are not systematic, but by rank $n=4$ there is
a systematic decrease in the error with decreasing cutoff.
Our results can be compared to those from the
separable expansions of low-momentum potentials considered
in Ref.~\cite{Kamada:2005hr}, which used polynomial expansions up to a
specified rank.  They found as well that calculations of the triton converge
to the exact result at lower rank as the cutoff is decreased,
with convergence to four digits or better at rank 10, 14 and 20 
for $\Lambda = 3, 5$ and $10 \fmi$, respectively.
Overall, the dramatic improvement in convergence of separable
expansions when cutoffs are lowered motivates a future study of how to
optimize such expansions. This study should also consider alternative
formulations, such as that of Ernst, Shakin and Thaler, which when
applied to conventional (and therefore high cutoff) potentials at
rank~4, leads to deviations of less than $50 \, \text{keV}$ for the 
triton binding energy \cite{SCHADOW00}.

\begin{figure}[t]
  \centerline{\includegraphics*[width=5.25in]{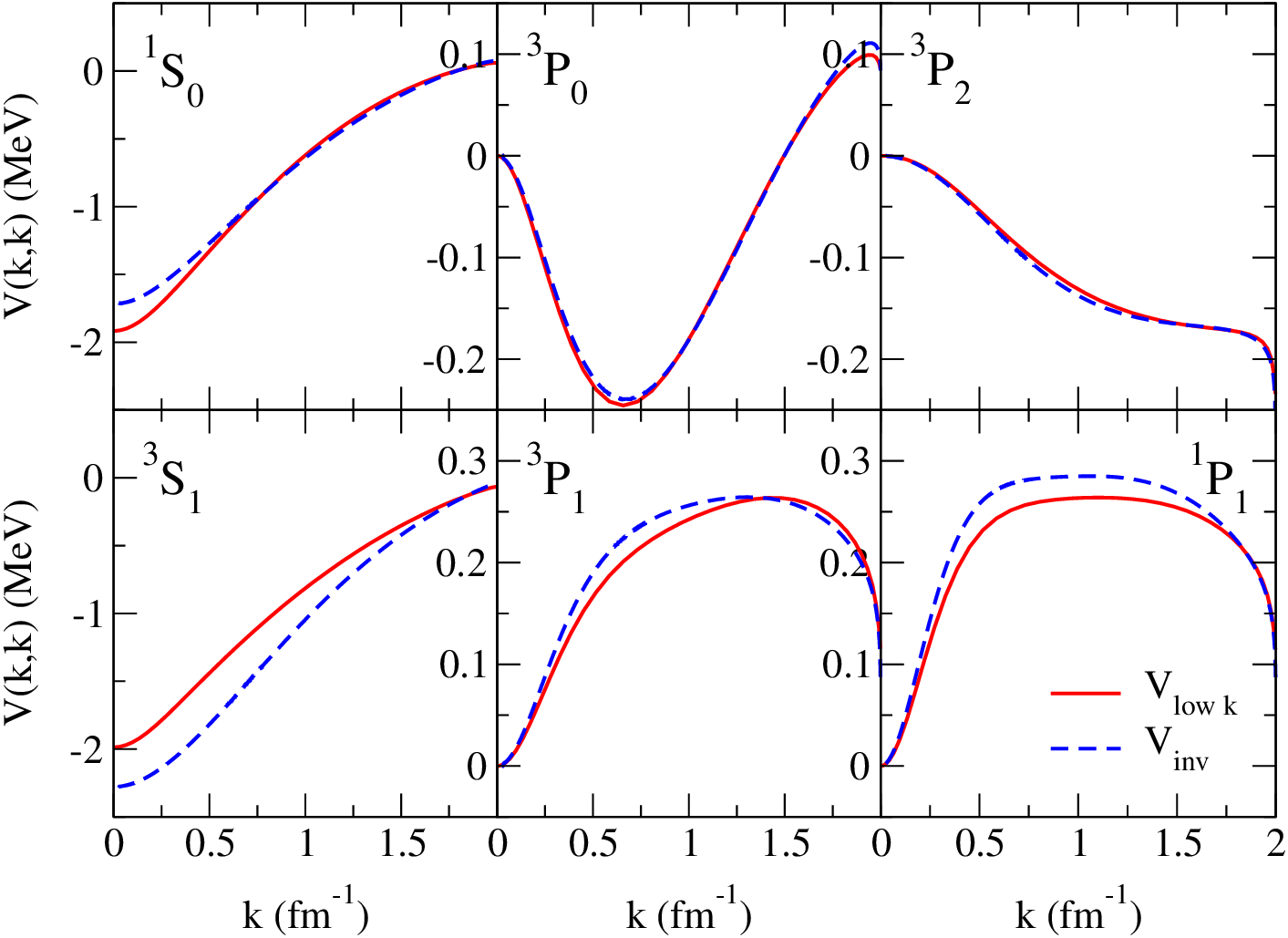}}
  \caption{Diagonal matrix elements $\vlowk(k,k)$ compared to separable
    inverse scattering results $V_{\rm inv}(k,k)$ for $\Lambda = 2.0
    \fmi$ in S and P-waves.}
  \label{fig:invscatt}
\end{figure}

In addition to the separable expansion in terms of Weinberg eigenvectors,
one can consider a separable inverse scattering 
approach~\cite{BrownJackson,Tabakin} that is based on
the scattering phase shifts and does not require a potential model. 
Here we make a simple 
exploratory study, using the separable inverse scattering solution
for a single channel to compare the inverse scattering low-momentum
interaction $V_{\rm inv}$ to $\vlowk$ in the S and P-waves. We therefore 
neglect the coupling of the $^3$S$_1$ and $^3$P$_2$ channels to the 
$^3$D$_1$ and $^3$F$_2$ waves respectively. The generalization 
to coupled channels and to higher rank separable interactions (such as 
an attractive and repulsive contribution) is straightforward~\cite{Tabakin}. 
For low cutoffs, the perturbative regime with Weinberg eigenvalues 
$|\eta| < 1$ suggests a direct connection to the phase shifts, except 
for the S-waves with near-zero-energy bound states. Our results show
that the inverse scattering solution may provide a good starting point 
to make this connection for separable interactions.

For a single channel, the solution to the inverse scattering problem
with a one-term separable interaction $V_{\rm inv}(k,k') = g \, v(k) 
\,v(k')$ is given by (where the sign of $g$ determines the sign of the 
interaction and $v(k) = \sqrt{V_{\rm inv}(k,k)/g}\,$)~\cite{BrownJackson,Tabakin}
\beqn
V_{\rm inv}(k,k) = - \biggl(\frac{k^2+k_{\rm b}^2}{k^2}\biggr) \,
\frac{\sin\delta(k)}{k} \: e^{-\Delta(k)} \,.
\label{eq:vinv}
\eeqn
The factor $(k^2+k_{\rm b}^2)/k^2$ is present for channels
with a bound state (in the $^3$S$_1$ partial wave $k_{\rm b} = \sqrt{m 
\, |E_{\rm d}|/\hbar^2}$ with deuteron binding energy $E_{\rm d}$) and
\beqn
\Delta(k) = \frac{2}{\pi} \, {\mathcal P}
\int\limits_0^\Lambda \frac{p \, dp}{p^2-k^2} \: \delta(p) \,.
\eeqn
Here we have generalized the conventional inverse scattering solution 
(where the phase shifts over \emph{all} energies are known) to an 
effective theory where all phase shifts are known up to the cutoff. 
Therefore $V_{\rm inv}(k,k)$ is defined only for momenta $k < \Lambda$.

In Fig.~\ref{fig:invscatt}, we compare the resulting inverse scattering
interaction $V_{\rm inv}(k,k)$ with $\vlowk(k,k)$ for a representative 
low-momentum cutoff $\Lambda = 2.0 \fmi$ in the S and P-waves. We find
that the separable inverse scattering solution reproduces qualitatively
all features of $\vlowk$. The largest differences are
observed in the S-waves, which is due to the rank-1 approximation. Note
that the (coupled channel) $^3$P$_2$ partial wave is well described in
our single-channel approximation, since the coupling to the $^3$F$_2$ wave
is weak. In addition, the inverse scattering solution leads to the same
characteristic dependences close to the cutoff, which makes it clear that
these are sharp cutoff artifacts. (Note that the separable inverse 
scattering solution would go smoothly to zero if one were to modify 
the phase shifts by a smooth cutoff regulator.) Finally, it is interesting
that the single-channel inverse scattering solution leads to a good
description of $\vlowk$ in channels where the interaction changes from 
attractive to repulsive, as is the case in the $^3$P$_0$ partial wave.
For low-momentum interactions, where the phase shifts are known over 
the range of validity of the effective theory, the inverse scattering 
approach therefore presents a promising addition to separable expansions
in Weinberg eigenvectors.

\section{Conclusions}
\label{sect:conclusions}

In this paper, Weinberg eigenvalues are used as a diagnostic to study
the nonperturbative nature of nucleon-nucleon interactions
as a function of a momentum cutoff.
As the cutoff is lowered using renormalization-group or model-space techniques,
the evolution of nonperturbative features at large cutoffs 
from strong short-range repulsion and iterated tensor interactions are
monitored via the repulsive complex Weinberg eigenvalues.
Major decreases in the magnitudes of repulsive eigenvalues
are observed for all conventional potentials when evolved to low momentum,  
even though two-body observables are unchanged.

When all eigenvalues for a given channel lie within the unit circle, 
the $T$ matrix expansion in terms of the potential $V$ is perturbative
in that channel, with the magnitude of the largest eigenvalues setting 
the rate of convergence. The repulsive eigenvalues satisfy this 
criterium in all channels for cutoffs of $2 \fmi$, where the crossover 
from the nonperturbative ($|\eta| > 1$) to the perturbative regime 
($|\eta| < 1$) happens for $\Lambda > 3\fmi$. The largest attractive 
eigenvalues, however, remain close to or beyond the unit circle in the 
S-waves, because they are due to the bound or nearly bound states.
Consequently, this is the only surviving source of nonperturbative
behavior for the Born series with low-momentum interactions. As shown 
in Ref.~\cite{Vlowknm}, the large attractive eigenvalues decrease 
rapidly due to Pauli blocking as the density is increased toward 
saturation density. This opens the possibility of perturbative and
therefore systematic calculations of nuclear matter. 

Chiral EFT interactions are also low-momentum interactions and
are expected to be soft. However, at N$^3$LO the isoscalar central part 
of the sub-sub-leading $2\pi$-exchange interaction is strongly
repulsive at short distances~\cite{Kaiser}. This leads to
large repulsive Weinberg eigenvalues for N$^3$LO potentials
in both singlet and triplet channels. As a result of these
nonperturbative features at N$^3$LO, the S-wave component
of the deuteron wave function~\cite{N3LOEGM} has a short-range 
part that is similar to what is found with a repulsive core.
In addition, the tensor force, e.g., from pion exchange, is 
singular at short distances. We have found major decreases
of the repulsive Weinberg eigenvalues as the cutoff is run
down from $\Lambda = 3 \fmi$ to $2 \fmi$, which tames the impact 
of the tensor force and of the nonperturbative features 
at N$^3$LO. This motives the evolution of chiral
interactions to lower cutoffs to improve their effectiveness
for few- and many-body problems~\cite{Vlowknm,Vlowksmooth}.

The decrease of the Weinberg eigenvalues
implies that separable expansions of nuclear interactions become
more effective for lower cutoffs. This was demonstrated for NN
scattering and the triton binding energy: for cutoffs of $2 \fmi$,
the typical relative error in the NN phase shifts is $0.1 \%$, and 
the convergence to the exact triton binding energy is to four digits 
at rank $6, 10, 12$ and $14$ for $\Lambda = 2, 3, 4$ and $5 \fmi$, 
respectively. Such separable expansions may be useful for four-body 
scattering~\cite{Fonseca,CSB} or to develop microscopic interactions
that can be handled by the current methods in nuclear reaction theory.
Perturbative Weinberg eigenvalues also suggest a closer connection 
of low-momentum interactions to phase shifts. This was explored
with rank-1 separable inverse scattering solutions, which were found 
to qualitatively reproduce the momentum dependences of $\vlowk$.

\begin{ack}
We thank Andreas Nogga for useful discussions, and Evgeny
Epelbaum for providing us with his code for the N$^2$LO and N$^3$LO 
potentials. This work was supported in part by the National Science 
Foundation under Grant No.~PHY--0354916 and the US Department of Energy 
under Grant No.~DE--FG02--97ER41014.
\end{ack}


\end{document}